\documentclass{ws-procs9x6}

\begin{document}

\title{Casimir stress in and force on a metal slab in a planar cavity}

\author{M. S. Toma\v s}
\address{Rudjer Bo\v{s}kovi\'{c} Institute, \\
P. O. Box 180, 10002 Zagreb, Croatia \\
%E-mail: tomas@thphys.irb.hr
}

\author{Z. Lenac}
\address{Department of Physics, University of Rijeka,\\
 51000 Rijeka, Croatia \\
%E-mail: zlenac@laboratory.com
}

\begin{abstract}
Emphasizing first the utility of the generalized Fresnel
coefficients in the theory of the Casimir effect in planar
cavities, we complement our previous discussion of the ordinary
Casimir force on and the Casimir stress in a metal (plasma) slab
in a planar cavity. We demonstrate strong dependence of the
Casimir stress in a thin slab on properties of the bounding medium
in the symmetric Lifshitz configuration. Contrary to this, the
stress in a thick slab gradually becomes insensitive on external
boundary conditions. We also consider the position dependence of
the Casimir force on and stress in a thin metal slab in a planar
cavity. Whereas the force per unit area on the slab strongly
increases when it approaches a mirror the stress in the slab
decreases and eventually changes the sign. Generally, the stress
decreases with the cavity width and decreasing reflectivity of the
mirrors.
\end{abstract}

\keywords{Casimir stress, metal slab, plasma model, planar
cavity.}

\bodymatter

\section{Introduction}
In addition to the ordinary Casimir forces \cite{Cas} acting
between the layers of a multilayered system, vacuum fluctuations
of the electromagnetic field cause a stress in each layer. This
(often disregarded) effect is important when considering
mechanical stability of thin layers and components\cite{BeCa} and
is therefore, besides being of fundamental interest\cite{Imry},
relevant in micro- and nano-technology. In our previous work
\cite{Braz} (referred to as I), we have considered the stress
(referred in that work as pressure) and its modal structure in a
metal (plasma) slab in the center of an ideal planar cavity and
demonstrated their strong dependence on the cavity width. Upon
emphasizing the utility of the concept of the generalized Fresnel
coefficients\cite{Craw} (in conjunction with their recurrence
relations\cite{TomG}) in the theory of the Casimir effect in
planar cavities\cite{Tom}, in this work we discuss the dependence
of the vacuum-field stress in a metal (plasma) layer in the
symmetric Lifshitz\cite{Lif} configuration on the properties of
the bounding medium. To emphasize the difference between the
standard Casimir force per unit area and the Casimir stress, we
also consider the position dependence of these quantities for a
metal slab in a planar cavity.

\section{Theory}
\begin{figure}
\begin{center}
\psfig{file=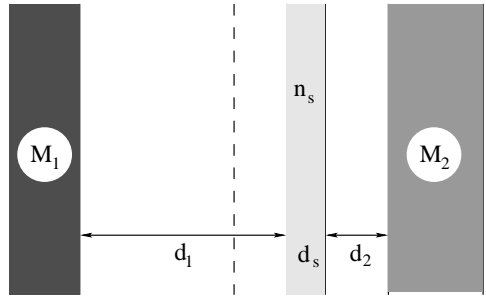,width=2.5in}
 \caption{\label{sys} A dielectric slab in na empty ($n_1=n_2=1$) planar
 cavity schematically. }
 \end{center}
\end{figure}

Consider a dielectric slab inserted in a planar cavity, as
depicted in Fig. \ref{sys}.
The vacuum-field forces (per unit area) acting on the slab consist
of the stress $F_s$ in the slab and the net slab-mirror
interaction force per unit area $F=F_2-F_1$ \cite{BeCa}, where
according to the theory of the Casimir force in multilayers
\cite{Tom}
\begin{equation}
\label{Fj} F_j={\rm
T^{(j)}_{zz}}=\frac{\hbar}{2\pi^2}\int_0^\infty d\xi \int^\infty_0
dkk\kappa_j\sum_{\rm TM,TE} \frac{r_{j-}r_{j+}e^{-2\kappa_jd_j}}
{1-r_{j-}r_{j+}e^{-2\kappa_jd_j}}.
\end{equation}
Here $\kappa_j(i\xi,k)=\sqrt{\varepsilon_j(i\xi)\xi^2/c^2+k^2}$ is
the perpendicular wave vector at the imaginary frequency in the
$j$th layer and $r_{j\pm}(i\xi,k)$ are the reflection coefficients
of the right and left stack of layers bounding the layer. $F$ can
be conveniently expressed in terms of the Fresnel coefficients $r
\equiv r_{1/2}=r_{2/1}$ and $t\equiv t_{1/2}=t_{2/1}$ of the {\it
whole} slab using the recurrence relation \cite{TomG,Tom}
\begin{equation}
r_{1+(2-)}(i\xi,k)=r+\frac{{t}^2R_{2(1)}e^{-2\kappa d_{2(1)}}}
{1-rR_{2(1)}e^{-2\kappa d_{2(1)}}},
\end{equation}
where $\kappa(i\xi,k)\equiv\kappa_1=\kappa_2=\sqrt{\xi^2/c^2+k^2}$
and $R_{1(2)}(i\xi,k)$ are reflection coefficients of the mirrors,
and noting that $r_{1-(2+)}=R_{1(2)}$. We find \cite{Tom}
\begin{eqnarray}
\label{F} F&=&\frac{\hbar}{2\pi^2}\int_0^\infty d\xi \int^\infty_0
dkk\kappa\sum_{\rm TM,TE}r \frac{R_2e^{-2\kappa
d_2}- R_1e^{-2\kappa d_1}}{N},\nonumber\\
N&=&1-r(R_1e^{-2\kappa d_1}+R_2e^{-2\kappa d_2})+
({t}^2-{r}^2)R_1R_2e^{-2\kappa (d_1+d_2)},
\end{eqnarray}
which agrees with the result obtained through a conventional way
\cite{Elli}.

Following Benassi and Calandra\cite{BeCa}, we ignore the
electostriction and magnetostrition forces in the slab. The stress
in the slab is then determined solely by the Minkowski stress
tensor \cite{Lan} and is therefore given by Eq. (\ref{Fj}), with
the reflection coefficients for the waves reflected within the
slab
\begin{equation}
\label{rs} r_{s-(+)}(i\xi,k)=\frac{-\rho+R_{1(2)}e^{-2\kappa
d_{1(2)}}} {1-\rho R_{1(2)}e^{-2\kappa d_{1(2)}}},
\end{equation}
where $\rho(i\xi,k)$ is the reflection coefficient of the
vacuum-slab interface.

\section{Discussion}
\begin{figure}
\begin{center}
\psfig{file=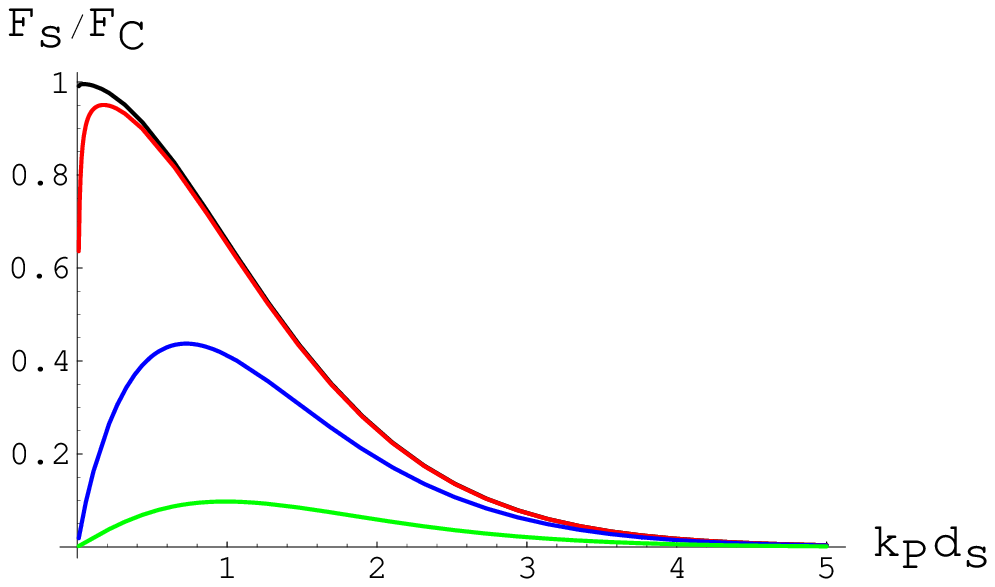,width=2in} \psfig{file=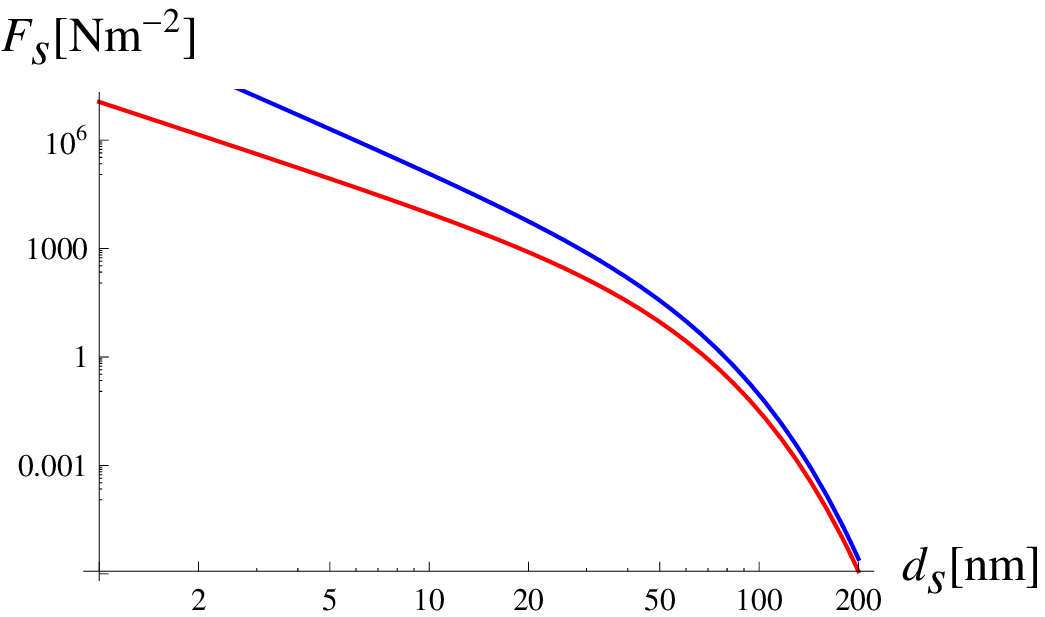,width=2in}
 \caption{\label{mirr} Left: Thickness dependence of the stress
 in a metal (plasma) layer. From top to bottom, the curves correspond to
 $\Omega_P/\omega_P=10^5, 10^3, 10$ and $1$, respectively.
 $k_P=\omega_P/c$ and $F_C=\pi^2\hbar c/240d_s^4$.
 Right: Stress in a gold layer between perfect ($\Omega_P=\infty$)
 mirrors (upper curve) and in a free-standing ($\Omega_P=0$) gold
 layer (lower curve) in absolute units.}
 \end{center}
\end{figure}

We first consider the stress in a metal layer sandwiched
($d_1=d_2=0$ in Eq. (\ref{rs})) between two identical (metal)
mirrors corresponding to the symmetric Lifshitz configuration.
Instead of a sophisticated model \cite{BeCa2,Esq}, we adopt here
the plasma model for the layer and the Drude model for mirrors
\begin{equation}
\varepsilon_s(i\xi)=1+\frac{\omega^2_P}{\xi^2},\;\;\;\;\;
\varepsilon_m(i\xi)=1+\frac{\Omega^2_P}{\xi^2+\Gamma^2},
\end{equation}
where $\omega_P$ and $\Omega_P$ are the corresponding plasma
frequencies and $\Gamma$ is the damping parameter of the mirrors
(in this work we use $\Gamma=10^{-3}\Omega_P$). The thickness
dependence of the stress is presented on the left side of Fig.
\ref{mirr} for several values of the (contrast) ratio
$\Omega_P/\omega_P$.
The uppermost curve practically coincides with the result obtained
assuming perfect ($\Omega_P=\infty$) mirrors. In that case the
stress in the layer can be calculated exactly and is, in the thin
layer ($k_Pd_s\ll 1$) limit, given by the Casimir force per unit
area $F_C$\cite{Braz}, which we used to scale the stress in the
figure. As seen, with decreasing reflectivity of the mirrors the
stress strongly drops and acquires in this range of the layer
thicknesses its nonretarded (nr) value $\sim d_s^{-3}$.
Ultimately, when $\Omega_P=0$, we obtain the stress in a
free-standing metal slab $F_s^{\rm
nr}=0.19k_Pd_sF_C$\cite{Imry,Braz}. Plotted on the right side of
Fig. \ref{mirr} is the stress in absolute units in a gold
($\omega_P=9$ eV \cite{BMM}) slab for two extreme cases of the
mirrors. Thus, the stress in an Au slab sandwiched between a
couple of realistic mirrors lies in between these two curves.

As noted already by Dzyaloshinski {\it et al}.\cite{Dzy}, the
stress in a thick metal (plasma) layer exponentially decreases.
Figure \ref{mirr} reveals, however, that for $k_Pd_s\gg 1$ it
becomes gradually insensitive to the properties of the mirrors.
For a thick enough layer, it is therefore given by the result
obtained for perfect mirrors\cite{Braz} $F_s=(\hbar
ck_P^4/4\sqrt{(\pi k_Pd_s)^3})\exp(-2k_pd_s)$. Since the large
$d_s$ behaviour of the stress is determined by small $\xi$ values
of $F_s(\xi,k)$, we note that the same conclusion applies to
layers described by a dielectric function of the form
$\tilde{\varepsilon}(i\xi)+\omega^2_P/\xi^2$, where
$\tilde{\varepsilon}(i\xi)$ behaves regularly at the origin. This
implies that addition of a salt into the liquid between the
plates, as in recent experiments on screened Casimir force
\cite{Mun}, will cause exponential decay of the force at large
liquid layer thicknesses since salt brings a plasma-like component
to the dielectric function of the solution.

\begin{figure}[h]
%\begin{center}
\psfig{file=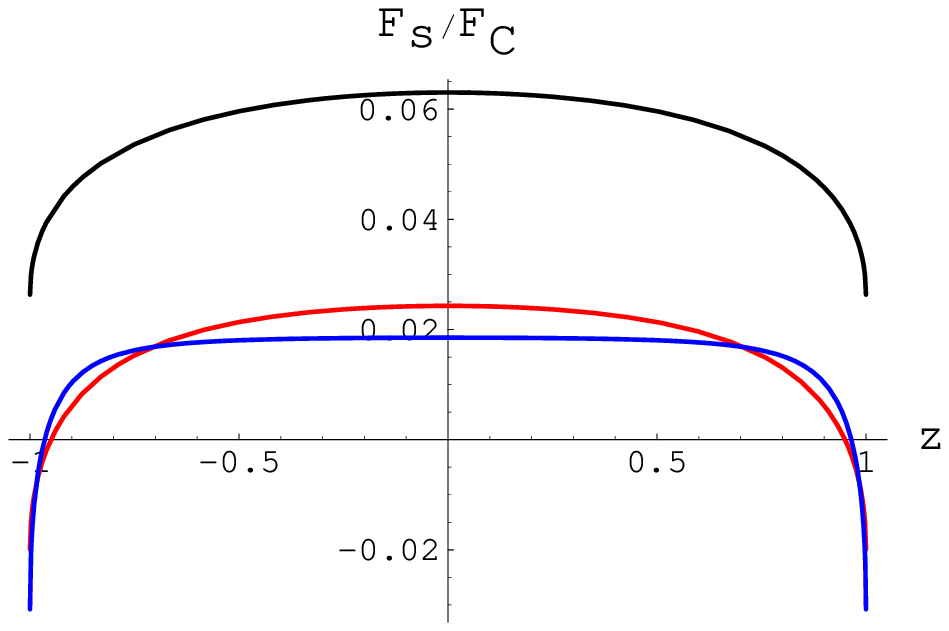,width=2in}
\psfig{file=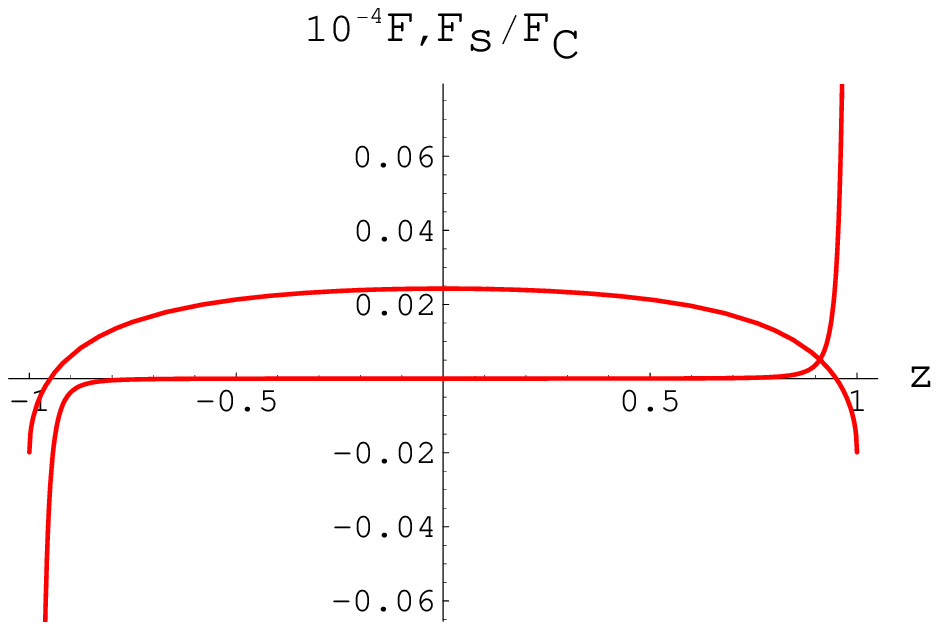,width=2in} \caption{\label{cav} Left:
Position dependence of the stress in a $k_pd_s=0.1$ thick  metal
slab in a cavity formed by mirrors with $\Omega_P=10^3\omega_P$.
From top to bottom, the curves correspond to the cavity width
$L=2d_s, 3d_s$ and $10d_s$, respectively. Right: Stress in vs net
force per unit area on the slab in the $L=3d_s$ cavity. Parameter
$z$ is defined with $d_{1(2)}=\frac{L-d_s}{2}(1\pm z)$. }
%\end{center}
\end{figure}

We end this discussion by briefly considering the stress in a
metal slab in a planar cavity. According to Eq. (\ref{rs}),
removing the mirrors from the slab ($d_1=d_2\neq 0$ in Eq.
(\ref{rs})) decreases its internal reflectivity. Accordingly, with
increasing slab-mirror distance, the stress in the slab behaves
similarly as in Fig. \ref{mirr} with decreasing reflectivity of
the mirrors (cf. with Fig. 4 of I). The position dependence of the
stress in a thin slab is illustrated in Fig. \ref{cav}.
As seen on left side of this figure, the stress is largest in the
center of the cavity (where the force vanishes) and decreases with
the cavity width until it saturates to $F^{\rm nr}_s$ (this
practically occurs already at $L=10d_s$). Near a mirror the stress
changes sign since $1<\varepsilon_s(i\xi)<\varepsilon_m(i\xi)$ is
fulfilled\cite{Dzy}. On the right side of Fig. \ref{cav}, we
compare the stress in and the force per unit area (scaled by a
factor of $10^{-4}$) on the slab in the $L=3d_s$ cavity. As
discussed by Benassi and Calandra\cite{BeCa}, these quantities
become approximately of the same order when the slab-mirror
distance is comparable with the slab thickness whereas at smaller
slab-mirror distances the Casimir force dominates.

%\section{Summary}
To summarize, in this work we have demonstrated strong thickness
and medium dependence of the Casimir stress in a metal layer in
the Lifshitz configuration as well as strong dependence of the
Casimir stress in and force on a thin metal slab in a planar
cavity on its position and cavity properties.

\section*{Acknowledgements}
This work was supported by the Ministry of Science, Education and
Sport of the Republic of Croatia under contract No.
098-1191458-2870.

\end{document}